\documentclass[preprints,conferenceproceedings,accept,pdftex,oneauthor]{Definitions/mdpi} 

\firstpage{1} 
\pubvolume{1}
\issuenum{1}
\articlenumber{0}
\pubyear{2023}
\copyrightyear{2023}
\datereceived{ } 
\daterevised{ } 
\dateaccepted{ } 
\datepublished{ } 
\hreflink{https://doi.org/} 
\pdfoutput=1 

\Title{Understanding perturbative and non-perturbative contributions to jets at RHIC}

\TitleCitation{Understanding perturbative and non-perturbative contributions to jets at RHIC}

\Author{Isaac Mooney $^{1,2}$\orcidA{}} 

\AuthorNames{Isaac Mooney}

\AuthorCitation{Mooney, I.}

\address{%
$^{1}$ \quad Wright Laboratory, Yale University, New Haven, CT; isaac.mooney@yale.edu\\
$^{2}$ \quad Brookhaven National Laboratory, Upton, NY\\}

\corres{Correspondence: isaac.mooney@yale.edu}

\abstract{Jets, as collections of multi-scale objects, allow for insight into perturbative (high-momentum) processes, but gaining an understanding of the non-perturbative structure within jets such as hadronization effects and the underlying event has been more difficult. In observables sensitive to these non-perturbative effects, a large discrepancy is observed between partonic calculations and experimental data at RHIC. Grooming techniques such as SoftDrop reduce the impact of these non-perturbative effects and isolate the hard radiation associated with the hard scattering for more direct comparison to perturbative calculations. By analogy, CollinearDrop can be used to isolate non-perturbative contributions from a different region of the emission phase space, or the Lund plane, although no measurement of a CollinearDrop observable has yet been published. In these proceedings, we review recent progress on experimentally accessing the perturbative and non-perturbative contributions to jets and their substructure, and discuss potential future measurements at RHIC as a road map toward precision QCD at the EIC.}
\usepackage{subcaption}
\begin{document}
\newcommand\pt{\ensuremath{p_{\mathrm{T}}}}
\section{Introduction}
Ultrarelativistic ion collisions typically produce many particles with low transverse momentum (\pt{}) with respect to the incoming beams. Rarely, however, high-momentum transdfers between quarks and gluons (\emph{partons}) in the incoming nucleons occur, leading to collimated groupings of particles with high total \pt{}, called jets. Jets are multi-scale objects which evolve from the initial high-momentum (\emph{hard}) scattered parton to lower energy quark and gluon daughters through successive radiation, until reaching virtualities near the hadronization scale $\Lambda_{\mathrm{QCD}}$, at which point partons recombine into hadrons. Perturbative quantum chromodynamics (pQCD) calculations are applicable for high energy processes, which correspond mostly to the parton shower component occurring early in jet evolution By contrast, soft or non-perturbative (np)QCD processes are generally not calculable from first principles. The technique of QCD on a spacetime lattice is an exception, although modeling dynamics is computationally intensive, and it is therefore inapplicable to jet studies. Partly due to difficulty theoretically describing the entirety of QCD jets, approaches were developed to avoid npQCD radiation's contribution to jet substructure observables. At the Relativistic Heavy Ion Collider (RHIC) the jet spectrum is softer than at the Large Hadron Collider (LHC), such that non-perturbative effects are more important. In Sec.~\ref{sec:pQCD} of these proceedings, we focus on recent measurements at RHIC using techniques which minimize the npQCD contribution. In Sec.~\ref{sec:npQCD}, we highlight measurements and observables which either improve our understanding of the npQCD contribution within jets, or which would benefit from our improved understanding of it. Finally, in Sec.~\ref{sec:discussion}, we give an outlook to future progress in the study of jets in their entirety.\par
\section{pQCD} \label{sec:pQCD} 
\begin{figure}[H]
\centering
\begin{subfigure}{0.32\textwidth}
	\includegraphics[width=\textwidth]{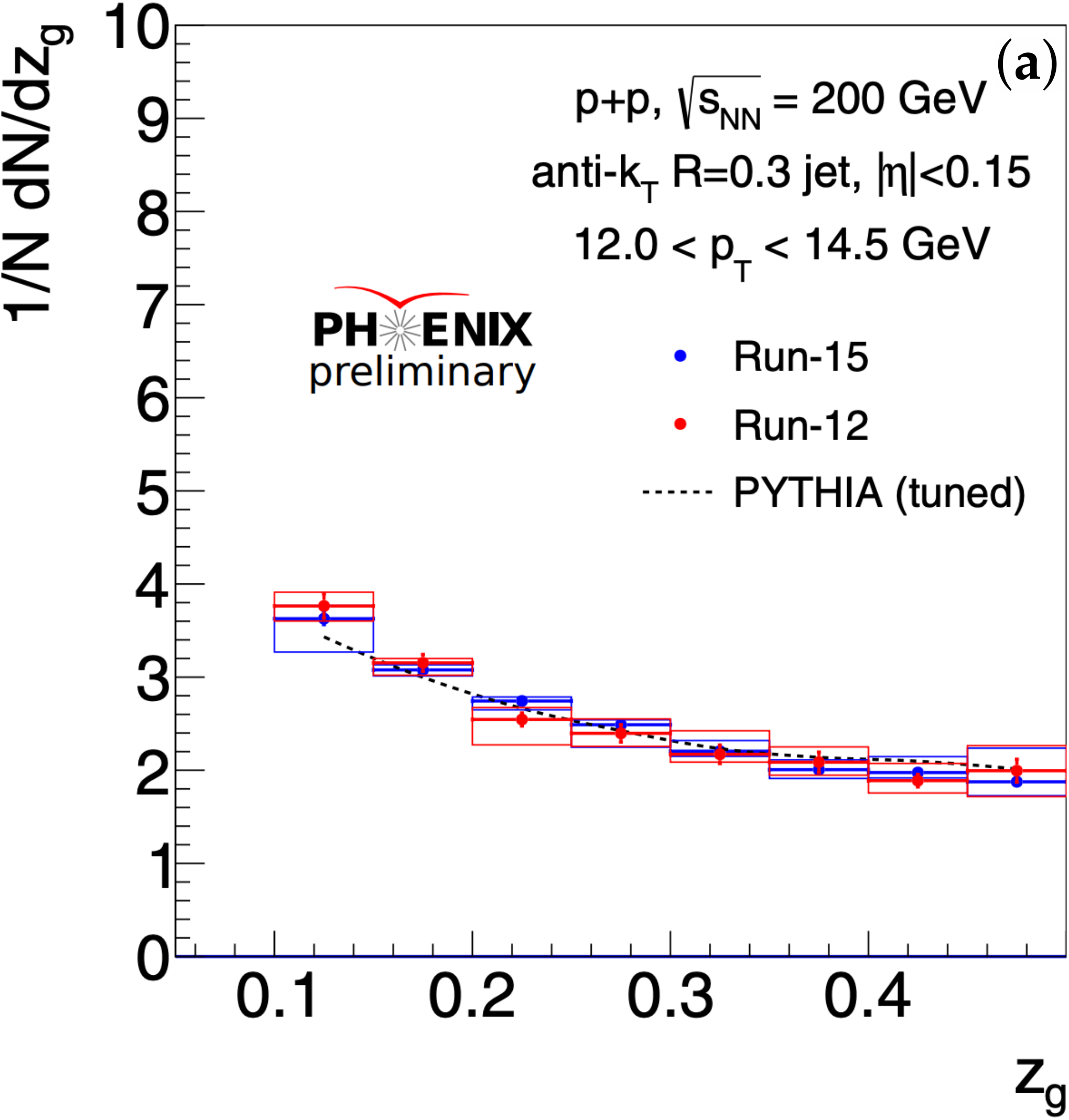}\phantomsubcaption{}\label{fig:SDphenix}
\end{subfigure}
\hfill
\begin{subfigure}{0.32\textwidth}
	\vfill
	\includegraphics[width=\textwidth, trim = 0cm -3cm 0cm 0cm]{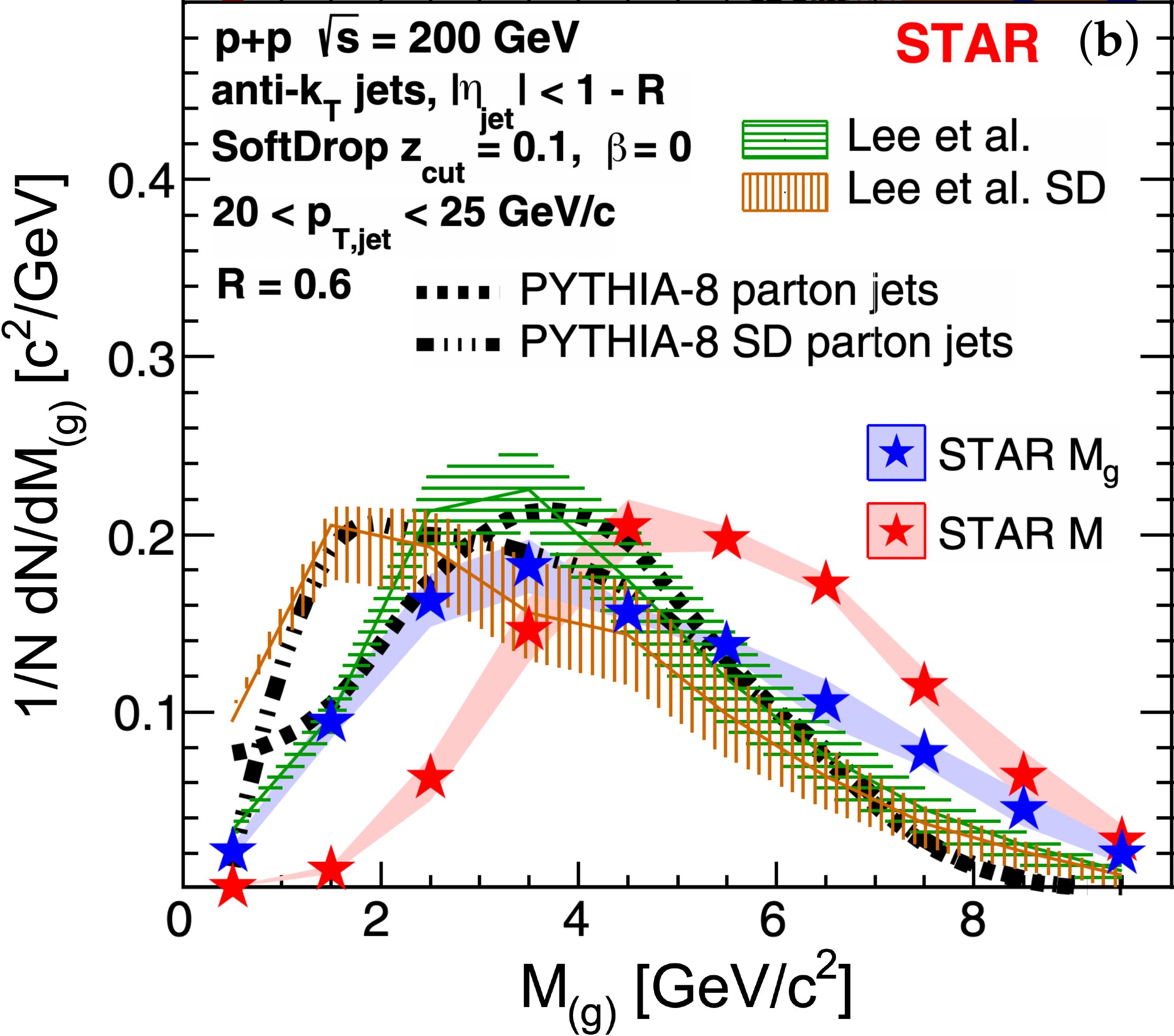}\phantomsubcaption{}\label{fig:SDstar}
	\vfill
\end{subfigure}
\hfill
\begin{subfigure}{0.32\textwidth}
	\includegraphics[width=\textwidth, trim = 0cm -1.2cm 0cm 0cm]{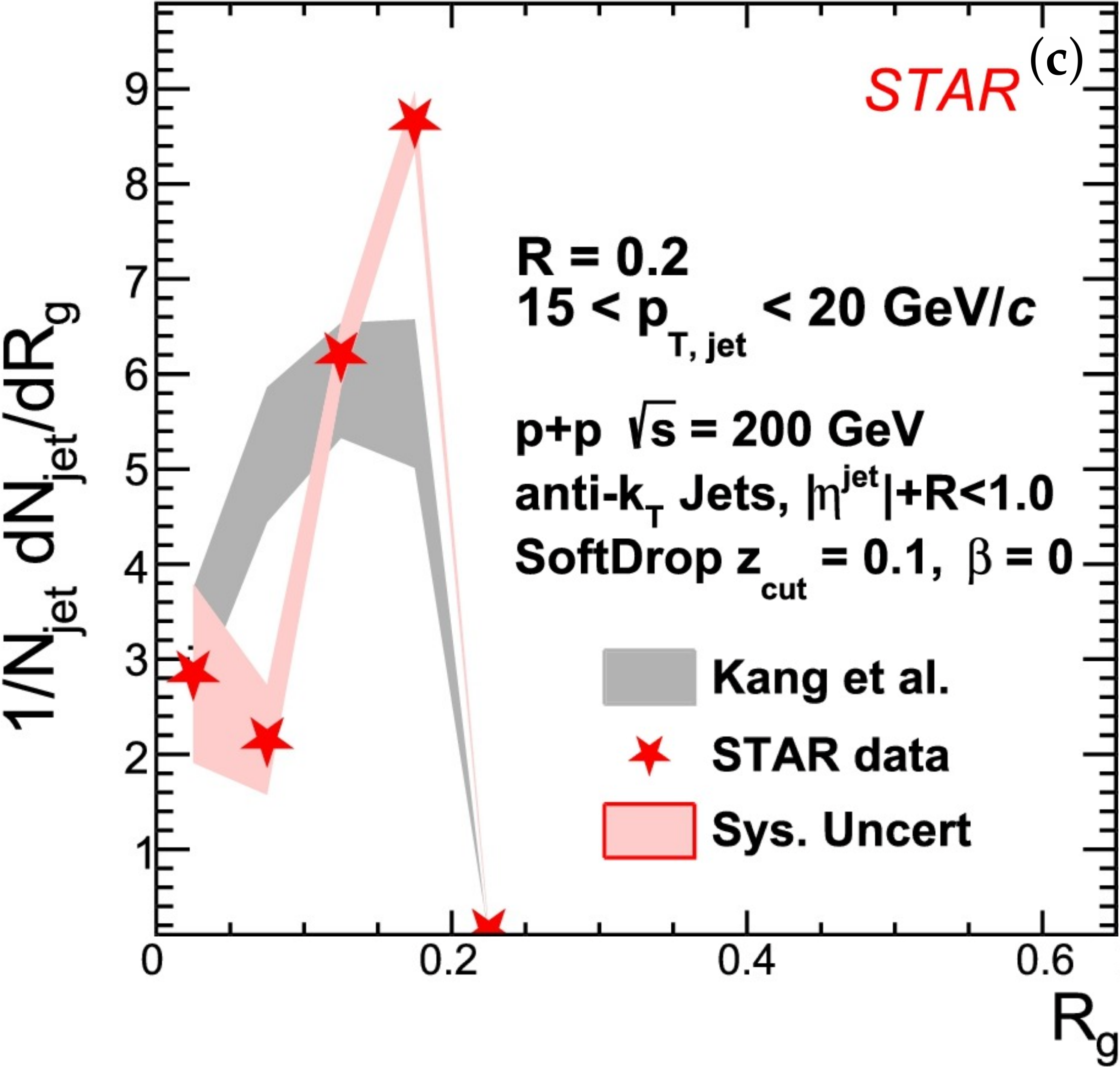}\phantomsubcaption{}\label{fig:SDstarmg}
\end{subfigure}
\caption{(\textbf{a}) Preliminary PHENIX measurement \cite{Belmont:2023afa} of momentum sharing fraction of the first hard split in the clustering history, $z_{\mathrm{g}}$. Data are compared to hadron-level MC event generator PYTHIA which demonstrates good agreement. (\textbf{b}) STAR measurement \cite{STAR:2021lvw} of the invariant mass (four-momentum sum of constituents) of the jet, $M_{(\mathrm{g})}$, before (red) and after (blue) SD grooming. Data are compared to a next-to-leading-logarithm accuracy (NLL) pQCD calculation (green) which qualitatively agrees with the groomed result for large jet \pt{} and $R$.  (\textbf{c}) STAR measurement \cite{STAR:2020ejj} of the opening angle of the first hard split, $R_{\mathrm{g}}$. Data are also compared to an NLL calculation, which again compares better at large jet \pt{} and $R$. Data in all cases are from $p$+$p$ collisions. \label{fig:SD}}
\end{figure}

The most commonly used algorithmic approach to reduction of soft, wide-angle radiation in a jet is called SoftDrop (SD) grooming \cite{Larkoski:2014wba}. Although the explicit goal is to minimize effects unrelated to a hard scattering's final state (e.g. pileup, initial state radiation, and underlying event), it is also very effective at reducing the non-perturbative contribution, including hadronization\nocite{Hoang:2019ceu}\endnote{There is still a small npQCD contribution to e.g. the groomed jet mass in the form of power corrections in the large mass region, with a larger contribution at smaller mass \cite{Hoang:2019ceu}.}. Recent results from proton-proton ($p$+$p$) collisions at RHIC have demonstrated good agreement with Monte Carlo (MC) event generators such as PYTHIA \cite{Bierlich:2022pfr} (Figure~\ref{fig:SDphenix}) and confirmed the assumption that SD-groomed observables are more directly comparable to perturbative, parton-level calculations (Figure~\ref{fig:SDstar}) although there remain discrepancies especially at small jet radii, $R$ (Figure~\ref{fig:SDstarmg}), since the physical scales of jet production and grooming go as $\pt R$ and $z_{\mathrm{cut}}\pt R$\endnote{$z_{\mathrm{cut}}$ is the momentum fraction threshold above which a given split in the jet shower is considered hard, and is not dropped.}, respectively, and these are closer to $\Lambda_{\mathrm{QCD}}$ for smaller radius.\par
Another way to reduce the non-perturbative contribution within jets is the subjet method \cite{Dai:2016hzf}. This involves reclustering a jet into subjets of smaller radius, $r < R$, at which point observables similar to those mentioned above (e.g. momentum sharing fraction, $z$, opening angle $R$ or $\theta$, etc.) can be constructed between subjets rather than between prongs. One advantage compared to similar studies of hadron momentum distributions within jets is that parton-to-hadron fragmentation functions, which are non-perturbative, are not involved in the theoretical calculations \cite{Kang:2017mda}. From an experimental perspective, they are also useful because they reduce background contamination in heavy-ion (A+A) collisions \cite{STAR:2021kjt}.\par
\section{npQCD} \label{sec:npQCD}
Although first-principles theoretical comparisons in the non-perturbative regime are often unavailable, this phase space region still deserves study. Widely used MC event generators such as PYTHIA include phenomenological modeling of hadronization and other npQCD processes, which requires tuning a parameter set to generate agreement with a large subset of collider data. PYTHIA 8.3's \cite{Bierlich:2022pfr} default tune, Monash \cite{Skands:2014pea}, did not include RHIC data which can create discrepancies with data even if the underlying perturbative physics mechanisms (matrix elements, parton shower, etc.) are modeled accurately. Specifically, subleading partonic interactions (MPI) contribute to the regularization of the partonic cross section at low-\pt, controlled in PYTHIA by a parameter called $p_{\mathrm{T},0}$ which enters the hard cross section as $\hat{\sigma} = (\pt^{2} + p_{\mathrm{T},0}^{2})^{-1}$.  The relation $p_{\mathrm{T},0} = p_{\mathrm{T},0}^{\mathrm{ref}}(\sqrt{s}/\sqrt{s}_{\mathrm{ref}})^{\mathrm{ecmPow}}$ is used to extrapolate from a reference center-of-mass energy, $\sqrt{s}$, to another, where $\mathrm{ecmPow}$ is a tunable parameter. Models tuned with LHC data may fail to describe RHIC data, over an order of magnitude lower in energy, due to this simplistic extrapolation. Adjusting $p_{\mathrm{T},0}^{\mathrm{ref}}$ and $\mathrm{ecmPow}$ among other MPI-related parameters as part of a new tune of PYTHIA 8.3 \cite{Aguilar:2021sfa} improved agreement with RHIC data, and with LHC data in some phase space regions.\par
\begin{figure}[H]
\centering
\includegraphics[width=0.35\textwidth]{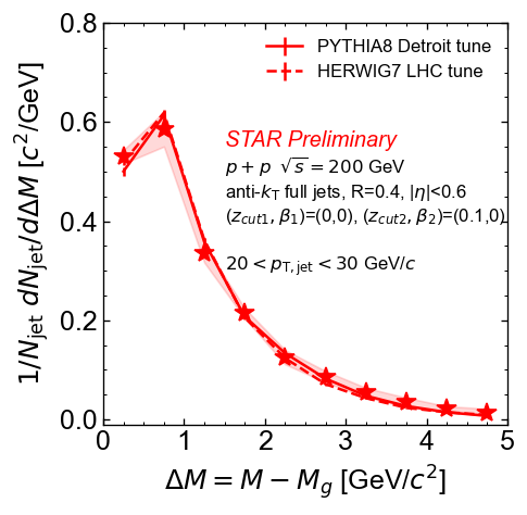}
\caption{Preliminary STAR measurement of CD jet mass in $p$+$p$ collisions. Data are compared to MC event generators PYTHIA and HERWIG which exhibit qualitative agreement with the data.}\label{fig:starCD}
\end{figure}

In addition to improving models such that current data can be described better, observables which increase sensitivity to soft radiation have also been proposed and investigated. One such example is CollinearDrop (CD) \cite{Chien:2019osu}, which can be considered roughly as the complement of SoftDrop, mentioned in Sec.~\ref{sec:pQCD}. Instead of grooming away the soft, wide-angle radiation, it is effectively the collinear radiation which is groomed away\endnote{A grooming of the softest, widest-angle radiation can also be performed to reduce e.g. the effect of pile-up.}. This anti-SD grooming results in a phase space with an enhanced non-perturbative contribution \cite{Stewart:2022ari}. Beneficially from an experimental point of view, it is also relatively simple to obtain a CD observable if the ungroomed and groomed observables have already been measured, as it is just the difference between the two. The STAR Collaboration has a recent preliminary measurement of the CD-groomed jet mass \cite{Song:2023sxb} (Figure~\ref{fig:starCD}).\par
Another way of accessing the soft phase space within jets is to examine subsequent branches in the splitting tree \cite{Frye:2017yrw}, beyond the first hard split as defined by SD. Since the jet is evolving with subsequent splittings toward $\Lambda_{\mathrm{QCD}}$, there is an expected larger npQCD contribution in later splits. In a STAR preliminary measurement , a behavior consistent with -- although not conclusively caused by -- this evolution was observed \cite{KunnawalkamElayavalli:2021ipx}.\par
Finally, in $p$+$p$ collisions, another promising avenue is the use of correlators of energy flow operators in the small angle limit, or more simply \emph{energy correlators} \cite{Basham:1978bw, Chen:2020vvp, Komiske:2022enw}. Because correlations of jet constituents separated by small (large) angular distance correspond to splittings occurring at late (early) times, the energy correlators exhibit an approximate separation between perturbative and non-perturbative regimes with no need to recluster the jets or remove npQCD contributions as in the case of the SD grooming procedure. STAR has recently made a preliminary measurement of the two-point energy correlator (EEC), which agrees well in the perturbative regime with an NLL pQCD prediction, and in the non-perturbative regime with a model assuming non-interacting hadrons \cite{Tamis:2023guc}.\par
In heavy-ion collisions, which produce a thermal medium called the quark-gluon plasma (QGP), often parton-medium interactions are calculated analytically in the weakly coupled limit. However, recent work (see e.g. Ref.~\cite{Moore:2021myb}) has started to incorporate multiple soft interactions with the medium into the calculation of the momentum broadening kernel. This provides a more realistic description of splitting rates within the medium, which in turn yields an accurate number of color sources for interaction with the medium. \par
\section{Discussion} \label{sec:discussion}
This diverse set of calculations, phenomenological modeling, and experimental results improves our understanding of the full structure of jets primarily by separating perturbative and non-perturbative effects more cleanly and enhancing the contribution of one or the other. Extensions to these works are ongoing and planned, and new avenues are being explored. For example, energy correlators may be measured at higher orders \cite{Komiske:2022enw}; in heavy-ion collisions \cite{Andres:2023xwr}; and separated by the constituents' charge \cite{Lee:2023npz}. The last point is part of an effort to examine specifically non-perturbative fragmentation into hadrons, and has motived measurement of charge correlators between the leading hadrons within jets \cite{Chien:2021yol}.
\vspace{6pt} 
\funding{This work was funded by the Office of Nuclear Physics of the U.S. Department of Energy under award number DE SC004168 and BNL/DOE-424803.}
\dataavailability{Most data mentioned above are available at \href{hepdata.net}{https://www.hepdata.net/}.}
\conflictsofinterest{The authors declare no conflict of interest. The funders had no role in the design of the study; in the collection, analyses, or interpretation of data; in the writing of the manuscript; or in the decision to publish the results.} 
\begin{adjustwidth}{-\extralength}{0cm}
\printendnotes[custom] 
\reftitle{References}
\bibliography{Mooney_Isaac}
\PublishersNote{}
\end{adjustwidth}
\end{document}